\documentclass[trans]{IEEEtran}
\usepackage{stfloats}
\usepackage{amsfonts}
\usepackage{amssymb}
\usepackage{amsthm}
\usepackage{cite}
\usepackage[cmex10]{amsmath}

\usepackage{float}
\usepackage{color}
\usepackage{algorithm}
\usepackage{algorithmic}
\usepackage{multirow}
\usepackage{graphicx}
\usepackage[normalem]{ulem}
\usepackage{tabularx}
\usepackage{subfigure}
\usepackage{amsmath}

\newtheorem{proposition}{Proposition}
\newtheorem{remark}{Remark}
\usepackage{fancybox,dashbox}
\usepackage{bbm}
\usepackage[justification=centering]{caption}
\usepackage[dvipsnames]{xcolor}
\usepackage{indentfirst}
\usepackage{mathtools}
\mathtoolsset{showonlyrefs}

\newcommand{\cmse}{\mathsf{CMSE}}
\newcommand{\myexpectation}[1]{\mathsf{E}{\left[#1\right]}}

\usepackage [margin=0.45in]{geometry}
\begin{document}
\title{Over-the-Air Computation via Broadband Channels}
	\vspace{-0.5cm}
\author{Tianrui Qin, Wanchun Liu$^\dagger$, Branka Vucetic and Yonghui Li
	\vspace{-1.2cm}
}
\maketitle
\begin{abstract}
\let\thefootnote\relax\footnote{The authors are with School of Electrical and Information Engineering, the University of Sydney, Australia.
	Emails:	\{tianrui.qin,\ wanchun.liu,\ branka.vucetic,\ yonghui.li\}@sydney.edu.au. (\emph{Wanchun Liu is the corresponding author.})
}Over-the-air computation (AirComp) has been recognized as a low-latency solution for wireless sensor data fusion, where multiple sensors send their measurement signals to a receiver simultaneously for computation.
Most existing work only considered performing AirComp over a single frequency channel.
However, for a sensor network with a massive number of nodes,
a single frequency channel may not be sufficient to accommodate the large number of sensors, and the AirComp performance will be very limited. 
So it is highly desirable to have more frequency channels for large-scale AirComp systems to benefit from multi-channel diversity.
In this letter, we propose an $M$-frequency AirComp system,
where each sensor selects a subset of the $M$ frequencies and broadcasts its signal over these channels under a certain power constraint. 
We derive the optimal sensors' transmission and receiver's signal processing methods separately, and develop an algorithm for joint design to achieve the best AirComp performance.
Numerical results show that increasing one frequency channel can improve the AirComp performance by threefold compared to the single-frequency case.
\end{abstract}


\section{Introduction}\label{intro}
During the past few years, over-the-air computation (AirComp) has emerged as a low-latency solution for large-scale wireless data fusion in many applications including Internet of Things~\cite{Goldenbaumletter,liu2019over,Zang20,Liu21,GuangxuMIMO,Wen19}, wireless distributed machine learning~\cite{Gunduz,Osvaldo,Broadband}, and over-the-air consensus~\cite{consensus}.
Normally, an AirComp system consists of multiple sensors and a fusion center computing the sum of the sensor measurements.
Multiple sensors send their measurement-carrying signals simultaneously to the fusion center over a multiple-access channel (MAC). 
The fusion center receives the superimposed signal of multiple sensors, and then estimates the sum of the sensors' measurements (see e.g.~\cite{liu2019over} for more details).

Many existing research focuses on optimal transmitter and receiver policy design of AirComp systems for achieving the minimum computation mean-square error (MSE) of the sum sensor signals~\cite{Goldenbaumletter,liu2019over,Zang20,Liu21,GuangxuMIMO,Wen19}.
In \cite{Goldenbaumletter}, the computation MSE under imperfect channel state information was investigated.
In \cite{liu2019over} and~\cite{Zang20}, 
the optimal transmitter (Tx) and receiver (Rx) scaling factor design of a single-antenna AirComp system under individual and sum power constraints were obtained, respectively.
{\color{black}In particular, a comprehensive
	comparison study of the AirComp design problem and
	the conventional minimum mean-square error (MMSE) problem
	was included in~\cite{liu2019over}, showing that the problem solutions are different.}
In~\cite{Liu21}, the optimal AirComp design with spatial and temporal correlated sensor signals was investigated.
In~\cite{GuangxuMIMO} and~\cite{Wen19}, the transmitting and receiving beamforming design problems of multi-antenna AirComp systems were considered. 

In addition to ~\cite{Goldenbaumletter,liu2019over,Zang20,Liu21,GuangxuMIMO,Wen19} that considers AirComp design over a single frequency channel, 
recently a broadband AirComp system with $M$ frequency channels was proposed in~\cite{Broadband} for distributed machine learning, where $M$ AirComp tasks were performed over $M$ frequency channels separately.
In other words, each frequency channel is dedicated to one AirComp task, which is the same as the single-frequency AirComp in~\cite{Goldenbaumletter,liu2019over,Zang20,Liu21,GuangxuMIMO,Wen19}.
For a large-scale sensor network, the performance of  AirComp over a single frequency is quite limited as many sensors may have poor conditions in that channel.
Thus, it is desirable to provide more frequency channels for large-scale AirComp systems to benefit from channel diversity.

In this work, we consider an $M$-frequency AirComp system and investigate the optimal Tx and Rx policy design.
Compared with the conventional single-frequency scenario, one needs to 1) determine the sensors' transmission policy at the $M$-frequency channel to optimally utilize the channel diversity for signal aggregation, noting that each sensor has access to any subset of the $M$ frequencies;
and 2) design the fusion center's policy to optimally recover the sum of the sensor signals based on the output of the $M$-frequency channel.
The system introduces an additional dimension for design compared to the single-frequency one and thus imposes the challenge.
%
To the best of our knowledge, such a problem has not been discussed in the open literature of AirComp.
The main contributions are summarized as follows.
\begin{itemize}
    \item We propose an $M$-frequency AirComp system, where each sensor decides to send its signal over a subset of the $M$ frequencies under a certain power constraint. The fusion center receives aggregated sensor signals at the parallel channels and then estimates the sum of the sensor signals.
   
   \item We formulate the computation MSE minimization problem of the $M$-frequency AirComp system, separately derive the optimal Tx and Rx policies in a closed-form, and adopt an alternating minimization approach for joint design. Our numerical results show that in the high-SNR scenario, the optimized two-frequency AirComp system can reduce the computation MSE to $1/3$ compared to the single-frequency case.

    \item After investigating the structure of the optimization problem, we obtain the property of the optimal AirComp system where the received sensor signals should be aligned onto a one-dimensional signal space at each complex (two-dimensional) frequency channel.
    By leveraging the property, we significantly reduce the computation complex of optimization.  
    Furthermore, our analysis show that the real number operation based fusion center performs better than the complex number operation based one in AirComp, though the latter is commonly adopted in the literature~\cite{liu2019over,Zang20,GuangxuMIMO,Wen19}.
%

\end{itemize}


\emph{Notations}: $\mathbb{R}$ and $\mathbb{C}$ denotes the set of real and complex numbers, respectively. $|\textbf{x}|$ denotes the Euclidean norm of vector $\textbf{x}$. $\myexpectation{\cdot}$ is the expectation operation. $\mathsf{Re}\{x\}$ and $\mathsf{Im}\{x\}$ are the real and imaginary parts of the scalar $x$.
\section{AirComp System over Multiple Frequency Channels}\label{system model}
\begin{figure}[t]  
	\centering  
	\includegraphics[width=0.5\textwidth]{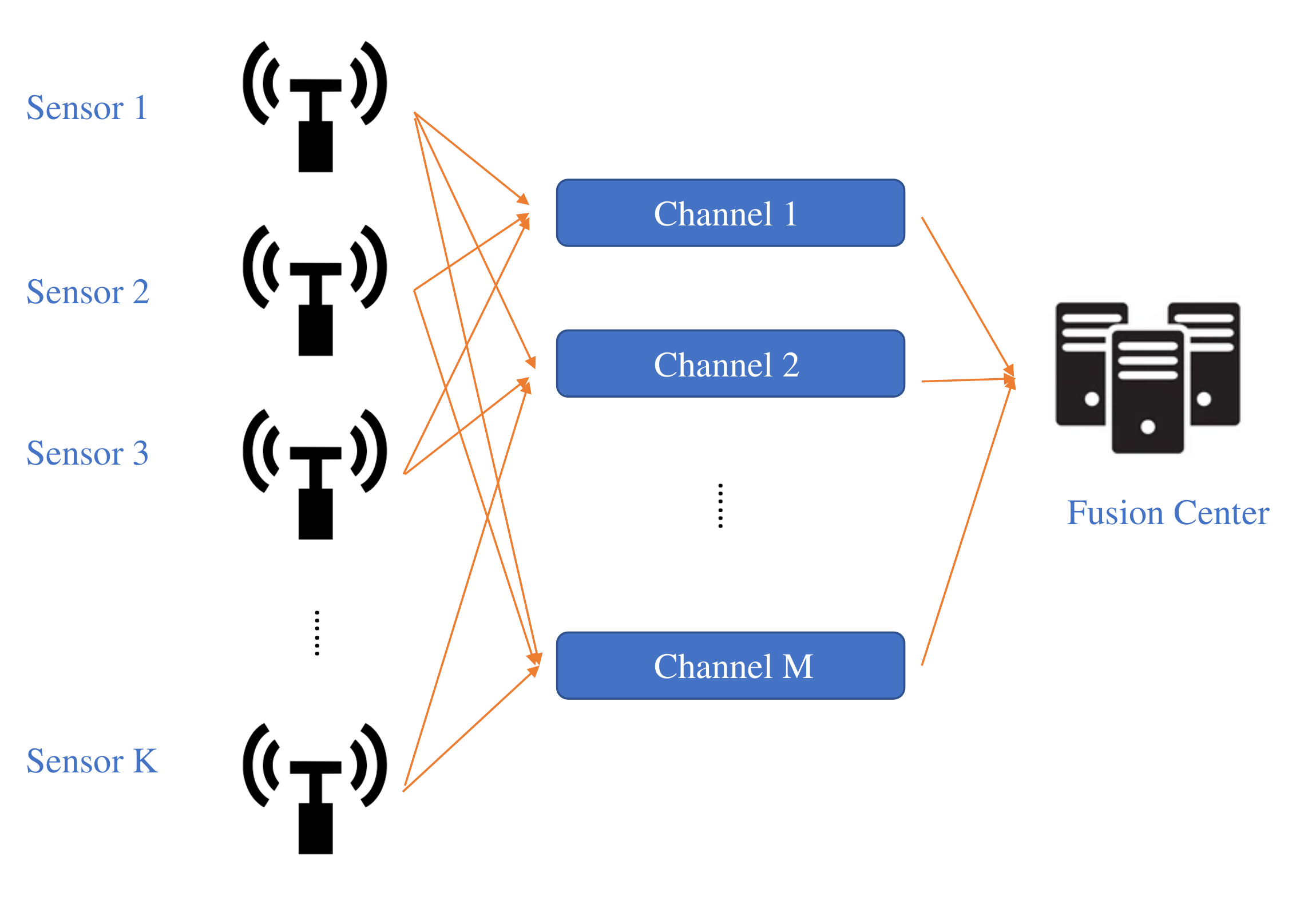}   
	\vspace{-0.5cm}	
	\caption{A $K$-sensor-$M$-frequency AirComp system.}    
	\label{system model fig}  
	\vspace{-0.5cm}
\end{figure}

We focus on a single-antenna AirComp system consisting of $K$ sensors, $M$ frequency channels (sub-carriers) and a fusion center (FC) as shown in Fig.~\ref{system model fig}, where $K\gg M$.
The channel coefficient between sensor $k$ and the FC at frequency $m$ is $h_{km} \in\mathbb{C}$.\footnote{\color{black}Note that the signaling cost for channel estimation grows linearly with the increasing number of frequencies.}
Sensor $k$'s measurement signal is $x_k\in \mathbb{R}, \forall k \in \mathcal{K}$, where $\mathcal{K} \triangleq \{1,2,\dots, K\}$. It is assumed that $x_k$ has zero mean and normalized variance~\cite{liu2019over,Zang20,Liu21,GuangxuMIMO}.\footnote{\color{black}Note that our AirComp framework of the scenario with zero-mean-normalized-variance sensor measurements can be applied to a generalized scenario with non-zero mean and non-normalized variance. Assume that sensor measurement ${x}_k$ has mean $\mu$ and variance $\delta^2, \forall k\in\mathcal{K}$.
The normalized measurement is $\bar{x}_k = \frac{x_k-\mu}{\delta}, \forall k\in\mathcal{K}$, and is sent to the fusion center based on the AirComp protocol.
Then, it is easy to have
$
\sum_{k=1}^{K} x_k =\delta \sum_{k=1}^{K} \bar{x}_k + K \mu.
$
Since $\delta$, $\mu$ and $K$ are constant, the optimal estimation of $\sum_{k=1}^{K} x_k$ can be obtained by the optimal estimation of  $\sum_{k=1}^{K} \bar{x}_k$ in the mean-square sense. Thus, in the following, we only consider the scenario with zero mean and normalized variance.}

In the multi-frequency AirComp system, each sensor can simultaneously broadcast its measurement via the $M$ frequency channels under a certain power constraint, and the FC estimates the sum signal $\sum_{k=1}^{K} x_k$ based on the received aggregated signal at the $M$ frequencies.
Let $b_{km}\in \mathbb{C}$ and $b_{km} x_k$ denote the Tx scaling factor and the transmitted signal of sensor $k$ at frequency $m$, respectively.
Certainly, $b_{km}=0$ indicates that sensor $k$ does not select frequency $m$ for AirComp. 
Assuming an average sensor power constraint $P$, we have
\begin{equation}\label{eq:constraint}
\sum_{m=1}^{M}  \myexpectation{\vert b_{km} x_k \vert^2} = \sum_{m=1}^{M} b^2_{km} \leq P, \forall k \in \mathcal{K}.
\end{equation}
The FC's received signal is 
\begin{equation}\label{eq:rm_1}
r_m=\sum_{k =1}^Kh_{km}b_{km}x_k+n_m, \forall m\in\mathcal{M}\triangleq \{1,2, \dots,M\},
\end{equation}
where $n_m\in\mathcal{CN}(0,\sigma^2)$ is the additive complex white Gaussian noise at frequency $m$.

As we consider complex channels, $r_m$ can be denoted as a two dimensional real vector $\textbf{r}_m \in \mathbb{R}^2$, and thus the FC receives $2 \times M$ real signals in total.
Then, \eqref{eq:rm_1} is rewritten as
\begin{equation}\label{channel output}
\textbf{r}_m=\sum_{k =1}^K\textbf{H}_{km}\textbf{b}_{km}x_k+\textbf{n}_m,
\end{equation}
where 
\begin{equation}\label{real_trans}
\begin{aligned}
    &\textbf{H}_{km}=\left[\begin{matrix}\mathsf{Re}\{h_{km}\} & -\mathsf{Im}\{h_{km}\}\\
                                    \mathsf{Im}\{h_{km}\} & \mathsf{Re}\{h_{km}\}\end{matrix}\right],\\
        &\textbf{b}_{km}=\left[\begin{matrix}\mathsf{Re}\{b_{km}\}\\\mathsf{Im}\{b_{km}\}\end{matrix}\right],\quad
        \textbf{n}_m=\left[\begin{matrix}\mathsf{Re}\{n_m\}\\\mathsf{Im}\{n_m\}\end{matrix}\right].
    \end{aligned}
\end{equation}
Note that $\mathsf{Re}\{n_m\}$ and $\mathsf{Re}\{n_m\}$ are real zero-mean Gaussian signals with variance $\sigma^2/2$.

Let $\textbf{g}_m \in \mathbb{R}^2$ denote the Rx scaling factor at frequency $m$.
The linear computation output of the FC is
\begin{equation}\label{received signal_c}
\begin{aligned}
        r=\sum_{m=1}^M\textbf{g}_m^\top\textbf{r}_m,
\end{aligned}
\end{equation}
and the computation MSE of the AirComp system is~\cite{GuangxuMIMO}
\begin{equation}\label{MSE}
    \cmse=\myexpectation{\Big|r-\sum_{k=1}^Kx_k\Big|^2}.
\end{equation}
{\color{black}We note that a smaller $\cmse$ indicates a better computation accuracy.}
Substituting \eqref{channel output} into \eqref{received signal_c} and then into \eqref{MSE}, we have
\begin{equation}\label{MSE_c}
\begin{aligned}
    \cmse&=\myexpectation{\Big\vert\sum_{k =1}^K(\sum_{m =1}^M\textbf{g}^\top_m\textbf{H}_{km}\textbf{b}_{km}-1)x_k+\sum_{m =1}^M\textbf{g}^\top_m\textbf{n}_m\Big\vert^2}\\
    &=\sum_{k =1}^K(\sum_{m =1}^M\textbf{g}^\top_m\textbf{H}_{km}\textbf{b}_{km}-1)^2+\frac{1}{2}\sigma^2\sum_{m=1}^M|\textbf{g}_m|^2.
\end{aligned}
\end{equation}

After some mathematical manipulations, \eqref{MSE_c} can be simplified as 
\begin{equation}\label{matrix MSE_c}
\begin{aligned}
\cmse=(\textbf{g}^\top\textbf{S} -\textbf{1}^\top)(\textbf{g}^\top\textbf{S} -\textbf{1}^\top)^\top+\frac{1}{2}\sigma^2
\textbf{g}^\top\textbf{g},
\end{aligned}
\end{equation}
where $\textbf{1}$ is a length-$K$ all-ones vector, 
$\textbf{g}^\top \triangleq \left[\textbf{g}_1^\top\  \textbf{g}_2^\top\  \dots \  \textbf{g}_M^\top\right]\in \mathbb{R}^{2M}$, 
and
\begin{equation}\label{eq:S}
   \textbf{S}= \left[\begin{matrix}
    \textbf{H}_{11}\textbf{b}_{11}&\textbf{H}_{21}\textbf{b}_{21}&...&\textbf{H}_{K1}\textbf{b}_{K1}\\
    \textbf{H}_{12}\textbf{b}_{12}&\textbf{H}_{22}\textbf{b}_{22}&...&\textbf{H}_{K2}\textbf{b}_{K2}\\
    ...&...&...&...\\
    \textbf{H}_{1M}\textbf{b}_{1M}&\textbf{H}_{2M}\textbf{b}_{2M}&...&\textbf{H}_{KM}\textbf{b}_{KM}
    \end{matrix}\right] \in \mathbb{R}^{2M\times K}.
\end{equation}

From \eqref{matrix MSE_c} and \eqref{eq:constraint}, the AirComp optimization problem is formulated as
\begin{equation}\label{eq:prob}
    \begin{aligned}
        &\min_{\textbf{g},\{\textbf{b}_{km}\}} ~~~(\textbf{g}^\top\textbf{S} -\textbf{1}^\top)(\textbf{g}^\top\textbf{S} -\textbf{1}^\top)^\top+\frac{1}{2}\sigma^2
    \textbf{g}^\top\textbf{g}\\
        &\text{~s.t.} ~~  \qquad \sum_{m=1}^{M}|\textbf{b}_{km}|^2\leq P,  \quad\forall k\in\mathcal{K}.
    \end{aligned}
\end{equation}

It is clear that the objective function of~\eqref{eq:prob} is non-convex due to the coupling of the Rx and Tx scaling factors $\textbf{g}$ and $\{\textbf{b}_{km}\}$, and there is no general approach to find the joint optimal solution. In what follows, we will \emph{separately} derive the optimal Rx and Tx scaling factors, and then adopt an alternating minimization approach for \emph{joint} design.

\section{AirComp Optimization}\label{optimal computation}
It is easy to prove that problem~\eqref{eq:prob} is bi-convex in terms of $\textbf{g}$ and $\{\textbf{b}_{km}\}$.
So we can alternately
solve~\eqref{eq:prob} for $\{\textbf{b}_{km}\}$ and $\textbf{g}$, respectively, while fixing the other, and then iterate until the convergence.
From~\cite{alternating} and references therein, the alternating minimization algorithm provides low-complexity solution of bi-convex problems. We will evaluate the effectiveness of the algorithm in solving the AirComp optimal design problem in Section~\ref{simulation}.
In the following, we will derive the optimal solution of the Rx and Tx scaling factors, respectively.

\subsection{Optimal Design of the Rx Scaling Factor}
Once the Tx scaling factors $\{\textbf{b}_{km}\}$ are given, \eqref{eq:prob} degrades to a quadratic problem without constraint.
The unique solution can be obtained by solving the equation below:
\begin{equation}\label{eq:derivative}
    \frac{\partial \cmse}{\partial \textbf{g}^\top}=2(\textbf{g}^\top\textbf{S} -\textbf{1}^\top)\textbf{S}^\top +\sigma^2\textbf{g}^\top = \textbf{0},
\end{equation}
which is simplified as
\begin{equation}
    \textbf{g}^\top(\textbf{S}\textbf{S}^\top+\frac{1}{2}\sigma^2\textbf{I})=\textbf{1}^\top \textbf{S}^\top.
\end{equation}

It is clear that $\textbf{S}\textbf{S}^\top+\frac{1}{2}\sigma^2\textbf{I}$ is positive definite and thus reversible. Then, we have the following result.
\begin{proposition}\label{pro:g}
	\normalfont
Given the Tx scaling factors $\{\textbf{b}_{km}\}$, {\color{black}the optimized Rx scaling factor} is 
\begin{equation}\label{give g}
    (\textbf{g}^\star)^\top=\textbf{1}^\top\textbf{S}^\top(\textbf{S}\textbf{S}^\top+\frac{1}{2}\sigma^2\textbf{I})^{-1},
\end{equation}
where $\textbf{S}$ is defined in~\eqref{eq:S}.
\end{proposition}
We see that a larger receiver noise leads to a smaller Rx scaling factor.

\subsection{Optimal Design of the Tx-Scaling Factors}
Once the Rx scaling factor $\textbf{g}$ is given, problem~\eqref{eq:prob} is decoupled into $K$ independent convex problems, where the $k$th problem is formulated as:
\begin{equation}\label{findbkm}
    \begin{aligned}
        & \min_{\textbf{b}_k} ~~~ L_k=\left(\sum_{m=1}^M\textbf{g}^\top_m\textbf{H}_{km}\textbf{b}_{km}-1\right)^2\\
        & \text{~~s.t.} ~~~  \vert \textbf{b}_k \vert^2-P\leq 0
    \end{aligned}
\end{equation}
where $\textbf{b}_k^\top \triangleq [\textbf{b}_{k1}^\top\  \textbf{b}_{k2}^\top\  \dots \ \textbf{b}_{kM}^\top ]$.
The Karush–Kuhn–Tucker (KKT) condition of problem~\eqref{findbkm} can be obtained as
\begin{equation}\label{KKT}
    \left\{
    \begin{array}{cc}
      \frac{\partial L_k}{\partial \textbf{b}_k} +\lambda \frac{\partial (\vert \textbf{b}_k \vert^2-P)}{\partial \textbf{b}_k}=\textbf{0} \\
         \lambda(\vert \textbf{b}_k \vert^2-P)=0 \\
       \vert \textbf{b}_k \vert^2-P\leq 0 \\
        \lambda \geq 0.
    \end{array}
    \right.
\end{equation}
After simplification, the first equation in~\eqref{KKT} is rewritten as
\begin{equation}\label{eq:opti_1}
(\textbf{h}_k\textbf{h}_k^\top+\lambda\textbf{I})\textbf{b}_k=\textbf{h}_k,
\end{equation}
where 
\begin{equation}\label{eq:hk}
\textbf{h}_k^\top= [(\textbf{H}_{k1}^\top \textbf{g}_1)^\top\ (\textbf{H}_{k2}^\top \textbf{g}_2)^\top\ \dots \ (\textbf{H}_{kM}^\top \textbf{g}_M)^\top\ ]. 
\end{equation}

To find the solution satisfying the KKT condition above, we consider the scenarios with $\lambda=0$ and $\lambda>0$.

1) If $\lambda=0$, \eqref{eq:opti_1} has infinite solutions as $\mathsf{Rank}(\textbf{h}_k\textbf{h}_k^\top) = \mathsf{Rank}([\textbf{h}_k\textbf{h}_k^\top, \textbf{h}_k]) = 1$, which is less than the dimension of $\textbf{h}_k\textbf{h}_k^\top$~\cite{anton2013elementary}.
By introducing the Moore-Penrose inverse operator $(\cdot)^+$~\cite[Corollary 1]{penrose_1955}, a general solution of~\eqref{eq:opti_1} can be written as
\begin{equation}\label{eq:w}
\textbf{b}_k=(\textbf{h}_k\textbf{h}_k^\top)^+\textbf{h}_k+(\textbf{I}-(\textbf{h}_k\textbf{h}_k^\top)^+\textbf{h}_k\textbf{h}_k^\top)\textbf{w},
\end{equation}
where $\textbf{w}$ can be any vector taken from $\mathbb{R}^{2M}$.

Next we will find the minimum-norm solution of $\textbf{b}_k$ in terms of $\textbf{w}$ and see if it satisfies the power constraint in~\eqref{KKT} or not.
We define a special solution to~\eqref{eq:opti_1} with $\textbf{w}=\textbf{0}$ in \eqref{eq:w} as
\begin{equation}
\textbf{b}^\circ_k=\textbf{A}^{+}\textbf{h}_k,
\end{equation}
where $\textbf{A} \triangleq \textbf{h}_k\textbf{h}_k^\top$.
From \eqref{eq:w}, we have
\begin{equation}\label{eq:norm_b}
\begin{aligned}
|\textbf{b}_k|^2&=|\textbf{b}_k-\textbf{b}^\circ_k|^2+2(\textbf{b}^\circ_k)^\top(\textbf{b}_k-\textbf{b}^\circ_k)+|\textbf{b}^\circ_k|^2.
\end{aligned}
\end{equation}
Using the properties of $\textbf{A}^+\textbf{A}\textbf{A}^+ = \textbf{A}^+$, $(\textbf{A}^+\textbf{A})^\top = \textbf{A}^+\textbf{A}$ and $\textbf{A}\textbf{b}_k = \textbf{h}_k$, it can be obtained that
\begin{equation}\label{eq:ls}
\begin{aligned}
(\textbf{b}^\circ_k)^\top(\textbf{b}_k-\textbf{b}^\circ_k)
&=(\textbf{A}^+\textbf{A}\textbf{b}^\circ_k)^\top(\textbf{b}_k-\textbf{A}^+\textbf{h}_k)\\
&=(\textbf{b}^\circ_k)^\top(\textbf{A}^+\textbf{h}_k-\textbf{A}^+\textbf{h}_k)=0.
\end{aligned}
\end{equation}
Taking \eqref{eq:ls} into \eqref{eq:norm_b}, it is straightforward to have
\begin{equation}
\begin{aligned}
|\textbf{b}_k|^2
=|\textbf{b}_k-\textbf{b}^\circ_k|^2+|\textbf{b}^\circ_k|^2
\geq|\textbf{b}^\circ_k|^2.
\end{aligned}
\end{equation}
Therefore, $\textbf{b}^\circ_k$ is the minimum-norm solution of $\textbf{b}_k$.
If $|\textbf{b}^\circ_k|^2 \leq P$, we have found an optimal solution of problem~\eqref{eq:prob}; otherwise, we consider the scenario below.

2) If $\lambda > 0$, $\textbf{h}_k\textbf{h}_k^\top+\lambda\textbf{I}$ is invertible and \eqref{eq:opti_1} has a unique solution:
\begin{equation}\label{eq:bk}
    \textbf{b}_k=(\textbf{h}_k\textbf{h}_k^\top+\lambda\textbf{I})^{-1}\textbf{h}_k.
\end{equation}

It is easy to see that the $2M \times 2M$ real symmetric matrix $\textbf{h}_k\textbf{h}_k^\top+\lambda\textbf{I}$ has all positive eigenvalues and increases with $\lambda$, and it has $2M$ linear independent eigenvectors that are invariant with $\lambda$.
Due to the fact that an inverted matrix has invariant eigenvectors but has reciprocal of the eigenvalues of the original matrix, it can be proved that $\vert \textbf{b}_k \vert$ of \eqref{eq:bk} monotonically decreases with $\lambda$.
Using the monotonicity, the solution of $\lambda$ satisfying the KKT condition can be easily obtained by solving the power constraint  $\vert \textbf{b}_k \vert^2 = P$ with a linear search approach. Then, the optimal $\textbf{b}_k$ can be obtained by \eqref{eq:bk}.

Therefore, the optimal solution of $\textbf{b}_k$ is summarized as below.
\begin{proposition}\label{pro:b}
\normalfont
Given the Rx scaling factor $\textbf{g}$, {\color{black}the optimized Tx scaling factor} $\textbf{b}_k,\forall k \in \mathcal{K}$, is 
\begin{equation}
\textbf{b}_k^\star = \begin{cases}
(\textbf{h}_k\textbf{h}_k^\top)^{+}\textbf{h}_k, &\text{ if } |(\textbf{h}_k\textbf{h}_k^\top)^{+}\textbf{h}_k|^2\leq P \\
(\textbf{h}_k\textbf{h}_k^\top+\lambda\textbf{I})^{-1}\textbf{h}_k, &\text{ otherwise}
\end{cases}
\end{equation}
where $\lambda$ is the unique solution of $|(\textbf{h}_k\textbf{h}_k^\top+\lambda\textbf{I})^{-1}\textbf{h}_k|^2= P$, and $\textbf{h}_k$ is defined in~\eqref{eq:hk}.
\end{proposition}
We see that the optimal Tx scaling factor $\textbf{b}_k^\star$ is a complex function of the channel coefficients $\textbf{H}_{k,m}, m \in \mathcal{M}$, and the Rx scaling factor $\textbf{g}$. Numerical results will be presented in Section~\ref{simulation} to show insights on power allocation in different frequency channels.

\section{Computational Complexity Reduction}\label{discussion}
In the previous section, an alternating minimization algorithm has been proposed to solve the AirComp design problem~\eqref{eq:prob} over multiple frequencies.
Next, we will further investigate the property of the objective function of \eqref{eq:prob} to reduce its computation complexity.

First, using the Cauchy–Schwarz inequality, we have
\begin{equation}\label{eq:equality}
\textbf{g}_m^\top\textbf{H}_{km}\textbf{b}_{km}\leq|\textbf{g}_m||\textbf{H}_{km}\textbf{b}_{km}|,
\end{equation}
where the equality holds once the angle between vectors $\textbf{g}_m$ and $\textbf{H}_{km}\textbf{b}_{km}$ is zero.
From the definition of $\textbf{H}_{km}$ in~\eqref{real_trans}, it is easy to show that the angle of vector $\textbf{H}_{km}\textbf{b}_{km}$ can be any value by properly designing $\textbf{b}_{km}$ without changing its norm.
So the equality \eqref{eq:equality} can hold for any given $\textbf{g}_m$ and $|\textbf{b}_{km}|$.

Second, observing the design target function~\eqref{MSE_c} and using the inequality~\eqref{eq:equality}, we have the inequality below with the same power constraint on $\textbf{b}_{k}$:
\begin{equation}\label{subs}
\min_{\{\textbf{b}_{km}\}}\!\!\left(\!\sum_{m=1}^M|\textbf{g}_m||\textbf{H}_{km}\textbf{b}_{km}|\!-\!1\!\!\right)^2 \!\!\!\leq \!\!
\min_{\{\textbf{b}_{km}\}}\!\!\left(\!\sum_{m=1}^M\textbf{g}_m^\top\textbf{H}_{km}\textbf{b}_{km}\!-\!1\!\!\right)^{\!2}\!\!.
\end{equation}
This is because the optimal design of the right hand side and the left hand side of~\eqref{subs} in terms of $\{\textbf{b}_{km}\}$ satisfy $\sum_{m=1}^M\textbf{g}_m^\top\textbf{H}_{km}\textbf{b}_{km}-1<0$ and $\sum_{m=1}^M|\textbf{g}_m||\textbf{H}_{km}\textbf{b}_{km}|-1<0$, respectively. Otherwise, one can always find smaller $\{|\textbf{b}_{km}|\}$ leading to smaller target functions in \eqref{subs}.
Then, from the discussions under~\eqref{eq:equality}, the equality in~\eqref{subs} holds if we design $\textbf{b}_{km}$ such that the angle between vectors $\textbf{g}_m$ and $\textbf{H}_{km}\textbf{b}_{km}$ is zero.

Last, taking \eqref{subs} into~\eqref{MSE_c}, the target function of problem~\eqref{eq:prob} is equivalent to
\begin{equation}\label{eq:cmse_real}
\cmse = \sum_{k =1}^K\left(\sum_{m =1}^M | \textbf{g}_m| |\textbf{H}_{km}\textbf{b}_{km}|-1\right)^2+\frac{1}{2}\sigma^2\sum_{m=1}^M|\textbf{g}_m|^2,
\end{equation}
which depends on the norms of vectors $\textbf{g}_m$ and $\{\textbf{H}_{km}\textbf{b}_{km}\}$ instead of the angles.
Thus, without loss of optimality, it is safe to assume that the second elements of $\textbf{g}_m$ and 
$\textbf{H}_{km}\textbf{b}_{km}$ are zero, i.e., 
\begin{equation}\label{eq:real_conv}
\begin{cases}
&\!\!\!\!\!\!\textbf{g}_m \triangleq \! [g_m,0]^\top,\ g_m \in \mathbb{R}\\
&\!\!\!\!\!\!\textbf{b}_{km} \triangleq \! 1/|h_{km}| \!\!\left[\begin{matrix}\mathsf{Re}\{h_{km}\} & \mathsf{Im}\{h_{km}\}\\
-\mathsf{Im}\{h_{km}\} & \mathsf{Re}\{h_{km}\}\end{matrix}\right]
\!\!\left[\begin{matrix}
b'_{km}\\0
\end{matrix}\right],\ b'_{km} \in \mathbb{R}.
\end{cases}
\end{equation}
In other words, only the real part of the received signal carries information in each frequency channel.
Therefore, in problem~\eqref{eq:prob}, by removing these zero elements, the sizes of $\textbf{g}$ and $\textbf{S}$ are reduced to half, which significantly reduces the computation complexity.
{\color{black}Based on the computational complexity of matrix operations~\cite{boyd2018introduction}, it is easy to obtain that the computational complexity of the optimized Rx and Tx scaling factors in Propositions~\ref{pro:g} and \ref{pro:b} are $\textit{O}(M^2K)$ and $\textit{O}(M^3)$, respectively. }

\begin{remark}\label{remark1}
For the optimal $M$-frequency AirComp system, the received signals from $K$ sensors are aligned onto a one-dimensional signal space at each (complex) frequency channel.
To fully utilize the two-dimensional complex signal space, one can perform two independent AirComp tasks of the $K$ sensors over the $M$ frequencies at the same time.
\end{remark}

\subsection{AirComp with Real or Complex Number Operations?}\label{comprehensive comp}
So far we have adopted the real number operation (RNO) on the received signals~\eqref{received signal_c}, treating real and imaginary part signals as individuals and linearly scaling each of them.
As mentioned earlier, many existing work of AirComp systems adopted complex number operation (CNO) on the received complex signal(s) at the FC~\cite{liu2019over,Zang20,GuangxuMIMO,Wen19}, treating the received complex signal as a wholes and linearly scaling it by a complex number.
Clearly, these two types of operations are different and the RNO is more powerful as it can process each part of the two-dimensional (complex) signal separately. In the following, we will discuss the CNO-based multi-frequency AirComp and compare it with the RNO-based one.

By adopting the CNO method, the linear computation output of the FC of the multi-frequency AirComp system~is
\begin{equation}\label{received signal_c_v2}
\begin{aligned}
r=\sum_{m=1}^M \tilde{g}_m {r}_m,
\end{aligned}
\end{equation}
where $\tilde{g}_m \in \mathbb{C}$ is the complex scaling factor at frequency $m$.
After simplification and by following the similar steps in Section~\ref{discussion}, the computation MSE defined in~\eqref{MSE} is equivalent to
\begin{equation}\label{eq:cmse_complex_2}
\widetilde{\cmse}=\sum_{k =1}^K\left(\sum_{m =1}^M |\tilde{g}_m|  |h_{km}b_{km}|-1\right)^2+\sigma^2\sum_{m =1}^M|\tilde{g}_m|^2.
\end{equation}
The only difference in the target function between RNO-based AirComp~\eqref{eq:cmse_real} and CNO-based one~\eqref{eq:cmse_complex_2} is the coefficient of noise-power term. In the RNO-based case, the noise power is halved, and thus leads to a smaller computation MSE and provides a $3$~dB SNR gain compared with the CNO-based case.
\begin{remark}
When adopting the CNO-based AirComp in~\eqref{received signal_c_v2}, both real and imaginary parts of the receiver noise contained in $r_m$ are involved in the computation. However, as discussed in Remark~\ref{remark1}, the optimal design should align the received signals onto a one-dimensional signal space per frequency channel. Therefore, the CNO-based AirComp has a higher computation MSE than the RNO-based one due to the effect of the additional noise, and the latter should be recommended.
\end{remark}

\vspace{-0.5cm}
\section{Numerical Results}\label{simulation}
In this section, we numerically evaluate the computation MSE of the $K$-sensor-$M$-frequency AirComp system with different $K$, $M$ and power constraint $P$, where the Rx and Tx scaling factors are obtained by the low-complexity alternating minimization algorithm in Sections~\ref{optimal computation} and \ref{discussion} with $20$ iteration rounds.
{\color{black}Note that our experiment results (not included due to the space limitation) show that the algorithm usually converges within 20 iteration rounds under a CMSE tolerance level of $10^{-3}$.}
Unless otherwise stated, we set $P=10$, the noise power $\sigma^2 = 1$~\cite{liu2019over,Zang20}, the number of sensors $K=50$.
We adopt normalized Rayleigh fading channels for evaluating the computation MSE that averages over $10^5$ random channel realizations.

\begin{figure}[t]  
	\centering  
	\includegraphics[width=0.5\textwidth]{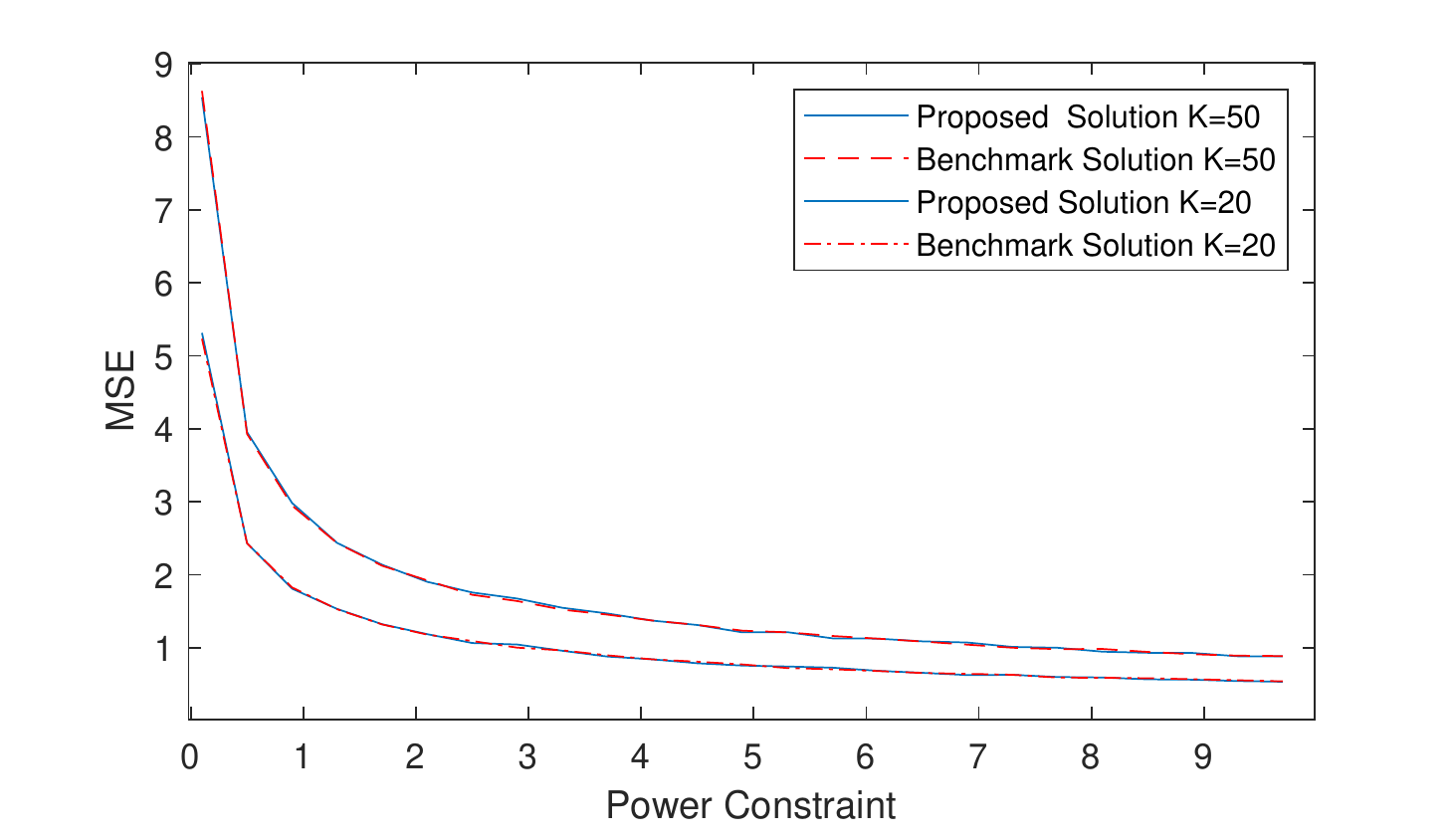}
	\vspace{-0.5cm}   
	\caption{Performance comparison between the proposed algorithm and the optimal one with $M=1$.}    
	\label{comparison_two_pairs}  
	\vspace{-0.5cm}
\end{figure}

In Fig.~\ref{comparison_two_pairs}, we demonstrate the effectiveness of the proposed alternating minimization method. Although the optimal joint Rx and Tx scaling factor design for a general $M$-frequency AirComp is unknown, the optimal single-frequency AirComp has been well investigated in~\cite{liu2019over}. We compare the AirComp performance achieved by the proposed algorithm and the benchmark optimal algorithm~\cite{liu2019over} for the single-frequency case with different $P$ and $K$.
We see that the proposed algorithm achieves exactly the optimal performance.

\begin{figure}[t]  
    \centering  
    \includegraphics[width=0.5\textwidth]{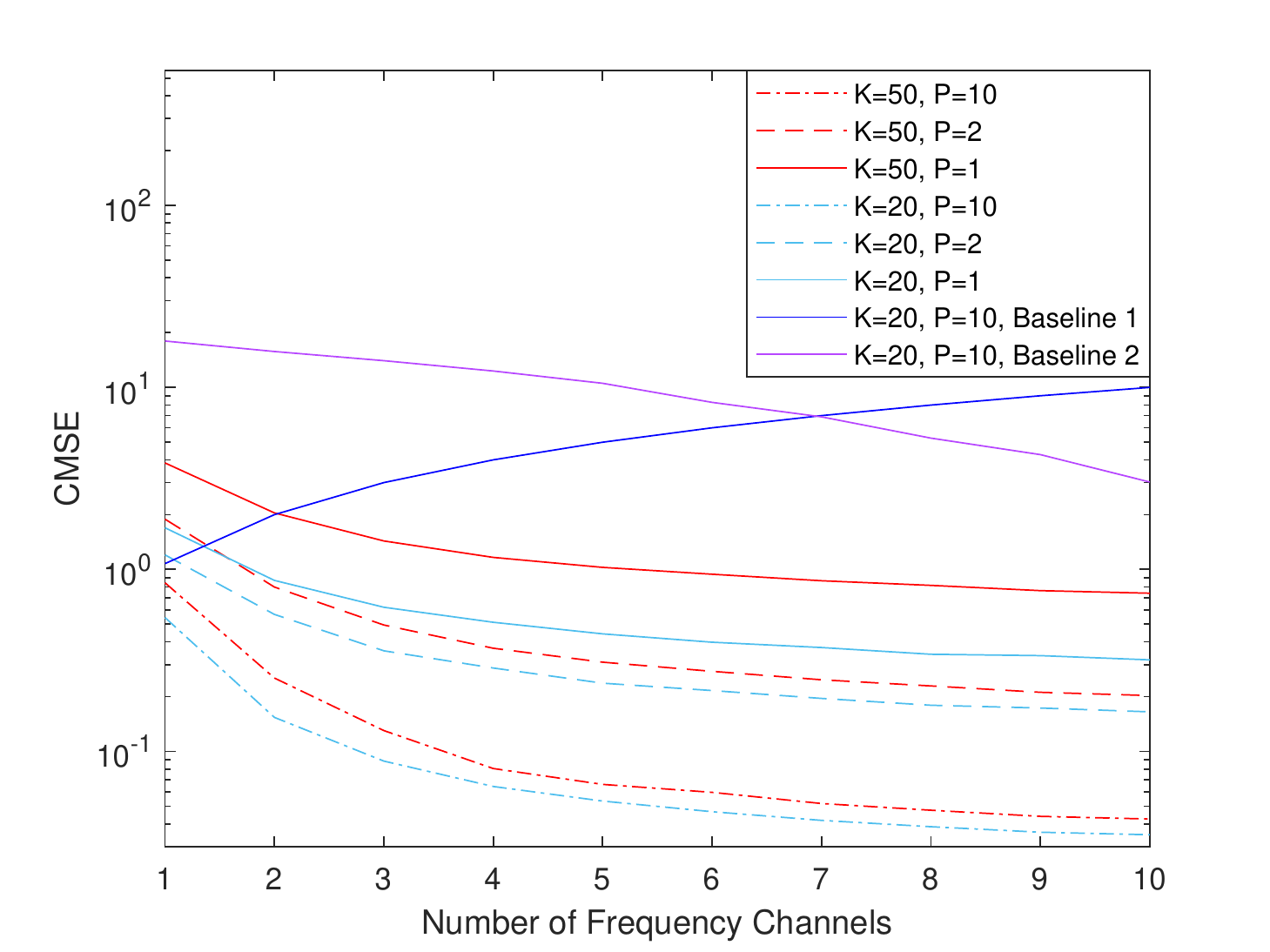}   
	\vspace{-0.5cm}       
    \caption{$\cmse$ versus $M$.}    
    \label{CMSE_M}  
	\vspace{-0.5cm}    
\end{figure}

In Fig.~\ref{CMSE_M}, we plot $\cmse$ with different number of frequency channels $M$. We see that $\cmse$ decreases with the increasing $M$ due to the increasing chance of having good channels for AirComp. It is interesting to see that when the transmit power is high (e.g., $P = 10$), increasing $M$ cannot reduce much $\cmse$ when $M>5$; while for the lower power scenario (e.g., $P =1$), increasing frequency channels is a good way to improve the AirComp performance.
In particular, when $P=10$, the optimized two-frequency AirComp system can reduce the computation MSE by $2/3$ compared to the conventional single-frequency case.
{\color{black}We also present $\cmse$ of two baseline policies: 1) the Rx scaling factors are fixed and are equal to one, and the Tx scaling factors are obtained by Proposition~\ref{pro:b}; 2) the Tx scaling factors are fixed and unified, i.e., the transmit power of sensor $k$ at frequency $m$ is equal to ${P/M}, \forall k\in \mathcal{K},m\in \mathcal{M}$, and the Rx scaling factors are obtained by Proposition~\ref{pro:g}.
It is clear that the proposed alternating minimization method provides significant performance gain compared to the the baseline policies.
In particular, baseline 1 leads to an increased CMSE when more frequencies are available, showing the importance of Rx scaling factor design.
}

\begin{figure}[t]  
    \centering  
    \includegraphics[width=0.5\textwidth]{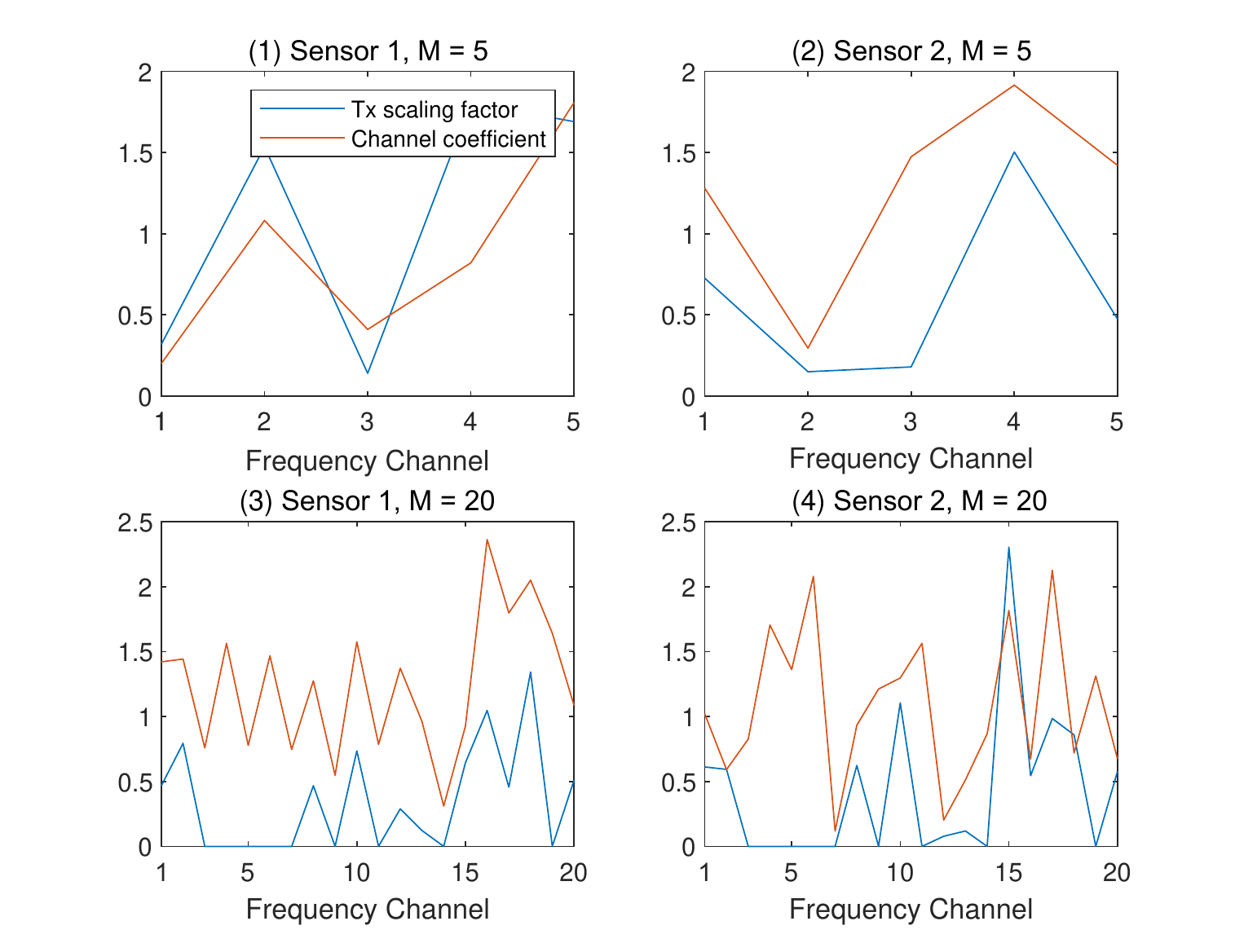}   
	\vspace{-0.7cm}       
    \caption{Channel coefficients and Tx scaling factors at different frequencies.}    
    \label{channel}  
	\vspace{-0.5cm}    
\end{figure}

In Fig.~\ref{channel}, we plot the optimized Tx scaling factors $\{|{b}_{km}|\}$ and the randomly generated channel coefficients $\{|h_{km}|\}$ at different frequencies for sensors 1 and 2 in sub-figures (1) and (2), respectively, with $M =5$, and in sub-figures (3) and (4) with $M=20$.
We see that for sensor 1, a larger channel coefficient leads to a higher Tx scaling factor in some frequencies, but it does not holds for some other cases. 
{\color{black}This is because as stated in Proposition~\ref{pro:b}, the optimized Tx scaling factors for one sensor depends on both its channel coefficients and the Rx scaling factors. Thus, the Tx scaling factors are not proportional to the channel coefficients.}
Comparing sub-figures (3) and (4) with (1) and (2), it is interesting to see that in the case with a large number of frequencies (e.g., $M= 20$), only half of the frequencies are utilized for AirComp as the allocated transmit power is zero in the other frequencies. It is because that the optimized AirComp system smartly chooses a subset of good channels for signal aggregation, instead of allocating transmit power to all frequency channels and suffering from more noise in a broader bandwidth.

\section{Conclusions}
We have proposed a novel $M$-frequency AirComp system, where each sensor broadcasts its signal over any a subset of the frequencies under a certain power constraint. 
We have derived the optimal Tx and Rx scaling factors separately in closed-form, adopted an alternating minimization approach for joint design, and further reduced the complexity of the design problem to half of it.
The numerical results have shown that in the high-SNR scenario, the optimized two-frequency AirComp system can reduce the computation MSE by $2/3$ compared to the conventional single-frequency case.
For future work, we will take into account the sensor signal correlation and the channel correlation at different frequencies, and study the optimal AirComp policy.


\bibliographystyle{IEEEtran}

\begin{thebibliography}{10}
	\providecommand{\url}[1]{#1}
	\csname url@samestyle\endcsname
	\providecommand{\newblock}{\relax}
	\providecommand{\bibinfo}[2]{#2}
	\providecommand{\BIBentrySTDinterwordspacing}{\spaceskip=0pt\relax}
	\providecommand{\BIBentryALTinterwordstretchfactor}{4}
	\providecommand{\BIBentryALTinterwordspacing}{\spaceskip=\fontdimen2\font plus
		\BIBentryALTinterwordstretchfactor\fontdimen3\font minus
		\fontdimen4\font\relax}
	\providecommand{\BIBforeignlanguage}[2]{{%
			\expandafter\ifx\csname l@#1\endcsname\relax
			\typeout{** WARNING: IEEEtran.bst: No hyphenation pattern has been}%
			\typeout{** loaded for the language `#1'. Using the pattern for}%
			\typeout{** the default language instead.}%
			\else
			\language=\csname l@#1\endcsname
			\fi
			#2}}
	\providecommand{\BIBdecl}{\relax}
	\BIBdecl
	
	\bibitem{Goldenbaumletter}
	M.~{Goldenbaum} and S.~{Stanczak}, ``On the channel estimation effort for
	analog computation over wireless multiple-access channels,'' \emph{IEEE
		Wireless Commun. Lett.}, vol.~3, no.~3, pp. 261--264, June 2014.
	
	\bibitem{liu2019over}
	W.~Liu, X.~Zang, Y.~Li, and B.~Vucetic, ``Over-the-air computation systems:
	Optimization, analysis and scaling laws,'' \emph{IEEE Trans. Wireless
		Commun.}, vol.~19, no.~8, pp. 5488--5502, Aug. 2020.
	
	\bibitem{Zang20}
	X.~{Zang}, W.~{Liu}, Y.~{Li}, and B.~{Vucetic}, ``Over-the-air computation
	systems: Optimal design with sum-power constraint,'' \emph{IEEE Wireless
		Commun. Lett.}, vol.~9, no.~9, pp. 1524--1528, Sep. 2020.
	
	\bibitem{Liu21}
	W.~{Liu}, X.~{Zang}, B.~{Vucetic}, and Y.~{Li}, ``Over-the-air computation with
	spatial-and-temporal correlated signals,'' \emph{Accepted by IEEE Wireless
		Commun. Lett.}, 2021.
	
	\bibitem{GuangxuMIMO}
	G.~{Zhu} and K.~{Huang}, ``{MIMO} over-the-air computation for high-mobility
	multimodal sensing,'' \emph{IEEE Internet Things J.}, vol.~6, no.~4, pp.
	6089--6103, Aug 2019.
	
	\bibitem{Wen19}
	D.~{Wen}, G.~{Zhu}, and K.~{Huang}, ``Reduced-dimension design of mimo
	over-the-air computing for data aggregation in clustered iot networks,''
	\emph{IEEE Trans. Wireless Commun.}, vol.~18, no.~11, pp. 5255--5268, 2019.
	
	\bibitem{Gunduz}
	M.~{Mohammadi Amiri} and D.~{Gündüz}, ``Machine learning at the wireless
	edge: Distributed stochastic gradient descent over-the-air,'' \emph{IEEE
		Trans. Signal Process.}, vol.~68, pp. 2155--2169, 2020.
	
	\bibitem{Osvaldo}
	J.~{Ahn}, O.~{Simeone}, and J.~{Kang}, ``Wireless federated distillation for
	distributed edge learning with heterogeneous data,'' in \emph{Proc. IEEE
		PIMRC}, 2019, pp. 1--6.
	
	\bibitem{Broadband}
	G.~{Zhu}, Y.~{Wang}, and K.~{Huang}, ``Broadband analog aggregation for
	low-latency federated edge learning,'' \emph{IEEE Trans. Wireless Commun.},
	vol.~19, no.~1, pp. 491--506, 2020.
	
	\bibitem{consensus}
	F.~Molinari, S.~Stanczak, and J.~Raisch, ``Exploiting the superposition
	property of wireless communication for average consensus problems in
	multi-agent systems,'' in \emph{Proc. ECC}, 2018, pp. 1766--1772.
	
	\bibitem{alternating}
	Q.~Li, Z.~Zhu, and G.~Tang, ``Alternating minimizations converge to
	second-order optimal solutions,'' in \emph{Proc. ICML}, 2019, pp. 3935--3943.
	
	\bibitem{anton2013elementary}
	H.~Anton and C.~Rorres, \emph{Elementary linear algebra: applications
		version}.\hskip 1em plus 0.5em minus 0.4em\relax John Wiley \& Sons, 2013.
	
	\bibitem{penrose_1955}
	R.~Penrose, ``A generalized inverse for matrices,'' \emph{Mathematical
		Proceedings of the Cambridge Philosophical Society}, vol.~51, no.~3, p.
	406–413, 1955.
	
	\bibitem{boyd2018introduction}
	S.~Boyd and L.~Vandenberghe, \emph{Introduction to applied linear algebra:
		vectors, matrices, and least squares}.\hskip 1em plus 0.5em minus 0.4em\relax
	Cambridge university press, 2018.
	
\end{thebibliography}
\end{document}